\documentclass{bvm} 
\usepackage{pgfplots}
\usepackage{cleveref}
\usepackage{comment}
\makeatletter
\let\blx@rerun@biber\relax
\makeatother
\begin{document}

\newcommand{\bvmyear}{2023}

\selectlanguage{english} 

\title{Attention-Guided Erasing: A Novel Augmentation Method for Enhancing Downstream Breast Density Classification}

\titlerunning{Attention-Guided Erasing: Breast Density Classification}




\author{
	Adarsh  \lname{Bhandary Panambur} \inst{1,2}, 
	Hui \lname{Yu} \inst{2}, 
    Sheethal \lname{ Bhat} \inst{1,2},
    Prathmesh \lname{Madhu} \inst{2}, 
	Siming \lname{Bayer} \inst{1,2},
	Andreas \lname{Maier} \inst{2}}

\authorrunning{Bhandary~Panambur et al.}

\institute{
\inst{1} Siemens Healthineers, Erlangen, Germany\\
\inst{2} Pattern Recognition Lab, FAU Erlangen-Nürnberg, Erlangen, Germany\\
}

\email{adarsh.bhandary.panambur@fau.de}



\maketitle

\begin{abstract}

The assessment of breast density is crucial in the context of breast cancer screening, especially in populations with a higher percentage of dense breast tissues. This study introduces a novel data augmentation technique termed Attention-Guided Erasing (AGE), devised to enhance the downstream classification of four distinct breast density categories in mammography following the BI-RADS recommendation in the Vietnamese cohort. The proposed method integrates supplementary information during transfer learning, utilizing visual attention maps derived from a vision transformer backbone trained using the self-supervised DINO method. These maps are utilized to erase background regions in the mammogram images, unveiling only the potential areas of dense breast tissues to the network. Through the incorporation of AGE during transfer learning with varying random probabilities, we consistently surpass classification performance compared to scenarios without AGE and the traditional random erasing transformation. We validate our methodology using the publicly available VinDr-Mammo dataset. Specifically, we attain a mean F1-score of 0.5910, outperforming values of 0.5594 and 0.5691 corresponding to scenarios without AGE and with random erasing (RE), respectively. This superiority is further substantiated by t-tests, revealing a \emph{p}-value of \emph{p}<0.0001, underscoring the statistical significance of our approach. 

\end{abstract}

\section{Introduction}
\label{intro}

With an estimated occurrence of 2.3 million cases worldwide each year, breast cancer stands as the most prevalent form of cancer among the adult population \cite{globocon}. The screening of patients for cancer is conducted through gold-standard mammography. Breast density, denoting the quantity of fibroglandular tissue in the breast, is known to be linked with the risk of breast cancer. Following the Breast Imaging Reporting and Database System (BI-RADS) scoring system \cite{birads}, breast density is categorized into four groups from A to D. Breast density A designates almost entirely fatty breast tissue, B corresponds to scattered fibroglandular tissues, C indicates heterogeneously dense tissue, and D denotes extremely dense tissue, as outlined by the BI-RADS classification \cite{birads}. In certain Asian populations such as Vietnam, women are recognized to exhibit dense breasts, a factor that is indicative of a higher risk of breast cancer \cite{trieu2017risk}. This may also influence the screening performance of radiologists, as higher breast density can also result in obscuring of underlying lesions in the mammography images, resulting in decreased mammographic sensitivity \cite{wang2014breast}. With the increase in breast cancer incidences worldwide and the population eligible for mammographic screening, automated methods can support streamlining the radiologists' workflow. The methods for automated breast density assessment primarily involve traditional volumetric methodology \cite{brandt2016comparison}, machine learning and deep learning-based approaches \cite{gardezi2019breast}. However, only a few research works have explored the task of automatically classifying breast density in populations with a very high percentage of individuals with extremely dense breasts, such as in Vietnam \cite{soat}. In this study, our goal is to classify breast density in mammogram images using a dataset from a Vietnamese screening population. Traditional deep learning methods typically employ separate networks to segment the background tissue or pectoralis muscle, especially in the case of mediolateral oblique (MLO) views at the pixel level, and subsequently classify the tissue into different density categories \cite{maghsoudi2021deep}. However, these methods rely on high-quality annotations provided either by radiologists or through traditional image processing techniques. Self-supervised learning (SSL) aims to learn robust lower-dimensional representations of images without using class information and has proven successful for downstream tasks such as classification. Recently proposed SSL techniques, leveraging vision transformers (ViT), have demonstrated robust localization capabilities in both natural and medical images \cite{dino, miao2022prior}. In this work, we propose a novel data augmentation method called Attention-Guided Erasing (AGE). AGE utilizes attention head visualizations extracted from a self-supervised ViT backbone trained using the DINO method \cite{dino}. These visualizations weakly localize the dense parts of the breast tissue. The generated maps are then employed to erase background regions, revealing only potential regions of dense breast tissues for the network to analyze. The proposed augmentation technique is specifically designed for use during the downstream transfer learning task of breast density classification.

The main contributions of our research are as follows: (a) We introduce a novel augmentation technique called Attention-Guided Erasing (AGE) for breast density classification in mammogram images, validated utilizing the VinDr-Mammo dataset \cite{vindr}. (b) We conduct extensive quantitative experiments, comparing the results with the traditionally used random erasing (RE) augmentation. Our findings demonstrate significant improvements when employing the AGE augmentation with varying random probabilities during training. (c) We present state-of-the-art results on the VinDr-Mammo test dataset \cite{vindr}. The paper is structured as follows: In Section \ref{materials}, a concise overview of the data and methods employed in our study is presented. Following that, Section \ref{results} showcases the quantitative outcomes for the breast density classification task and also offers a brief analysis and discussion of the results.

\begin{table}[t]
\begin{tabular*}{\textwidth}{l@{\extracolsep\fill}ccccc}

\hline                          
         & Breast Density A & Breast Density B  & Breast Density C  & Breast Density D \\  \hline
Training    & 72        & 1308           & 10366         & 1854     \\
Validation  & 8         & 220            & 1866          & 306    \\   
Test        & 20        & 380            & 3060          & 540    \\       
Total       & 100       & 1908           & 15292         & 2700     \\ \hline
\end{tabular*}
\caption{Class sample distribution in VinDr-Mammo dataset for breast density classification~\cite{vindr}.}
\label{data_table}
\end{table}

\section{Materials and Methods}
\label{materials}

\subsection{Data} 
\label{data}

We utilize the publicly available VinDr-Mammo dataset, which comprises full-field digital mammography images obtained from Hospital 108 and Hanoi Medical University Hospital in Hanoi, Vietnam \cite{vindr}. The dataset includes 20,000 images acquired from 5,000 patients, encompassing both craniocaudal (CC) and MLO views from both the right and left breasts of each patient. Three experienced radiologists reached a consensus in analyzing the dataset to annotate the BI-RADS and breast density \cite{birads}. The distribution of class samples utilized in this work is depicted in \cref{data_table}. We employed the original distribution of the training set released by the authors of VinDr-Mammo \cite{vindr} to split the training and validation datasets. The split is conducted at the patient level, and additional care is taken to ensure an equal distribution of class samples, along with considering the distribution of multiple mammography systems used in the data acquisition. Additionally, we reused the original test dataset. In total, the training, validation, and test datasets consist of 13,600, 2,400, and 4,000 images, respectively.

\begin{figure}[t]
\centering
\includegraphics[width=0.80\textwidth]{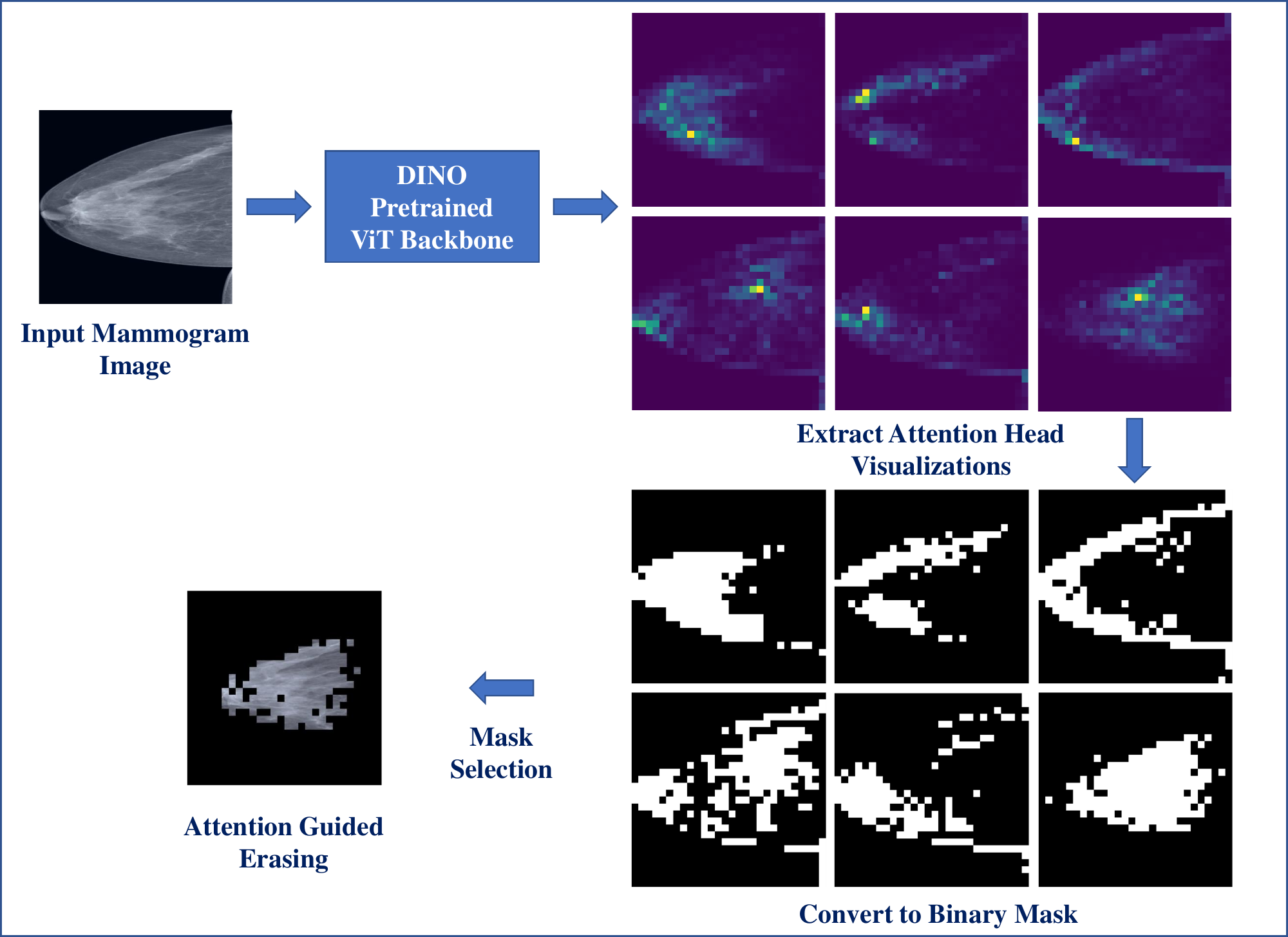}
\caption{Overview of the proposed methodology for the Attention-Guided Erasing (AGE).}  
\label{methodology}
\end{figure}

\vspace{0cm}
\subsection{Methodology}
\label{ssl-dino}

Figure \ref{methodology} shows an overview of the proposed augmentation methodology. We use the training dataset to train a ViT based on the DINO self-supervised pretraining method \cite{dino}. Subsequently, we use the trained backbone to extract visualizations from the six attention heads of the pretrained backbone. We select the best attention head that potentially localizes dense breast tissue and then erase the background regions around the weakly localized breast tissue. This process is intended to encourage the network to focus more on the dense tissue than the surrounding regions in the breast during transfer learning. In order to train a robust backbone with localization capability for our classification network, we perform self-supervised pretraining on the VinDR-Mammo training dataset using the DINO method~\cite{dino}. DINO employs a knowledge distillation approach where a student network learns to predict the teacher network outputs by utilizing a cross-entropy loss to match the softmax probabilities of the student and teacher outputs \cite{dino}. Two views consisting of different random transformations of the input image are fed into the same architecture-based student and teacher networks with different parameters \cite{dino}. The exponential moving average of the student network's parameters is used to update the teacher network to avoid the model collapse \cite{dino}. We reuse the implementation as suggested in the original work in this stage, including the ImageNet pretrained DeiT-small ViT and the standard set of randomly applied data augmentations including horizontal flip, colour jitter, gaussian blur, and solarization ~\cite{dino}. For input preparation, a random resized crop was applied at both global (0.4, 1.0) and local scales (0.05, 0.4) to the 224$\times$224 resolution of the full-resolution mammogram image, with a patch size of 16 serving as the input to the network~\cite{dino}. During pretraining, we employed a modest batch size of 32 to conform to the limitations of single GPU memory, leveraging a Quadro RTX 5000. The networks were trained for 300 epochs to facilitate thorough learning, with the final model being saved at the point of least training loss~\cite{dino}. 

\begin{table}[t]
    \centering
    \setlength{\tabcolsep}{8pt}  
    \begin{tabular*}{0.5\textwidth}{c @{\extracolsep{\fill}} c}
        \hline
        Model & Macro F1-Score  \\ 
        \hline
        ResNet-34 \cite{soat} & 0.504 \\
        EfficientNet-B2 \cite{soat} & 0.5525 \\
        DINO (No Erasing) & 0.5594 (0.026) \\
        DINO (RE, \emph{P} = 0.2) & 0.5691 (0.020) \\
        \textbf{DINO (AGE, \emph{P} = 0.6)} & \textbf{0.5910} \textbf{(0.017)} \\
        \hline
    \end{tabular*}
    \caption{Classification performance on the VinDr-Mammo test dataset. \emph{P} indicates the probability of erasing augmentations on the inputs during training. The best results are in \textbf{bold}.}
    \label{tab:scores}
\end{table}

After training the model in a self-supervised manner, we analyze the self-attention exhibited by six attention heads focused on the [CLS] token in the final layer. Visual inspection indicated that each attention head concentrated on a distinct region of the image. By analyzing the activation patterns of these attention heads, we were able to weakly identify the one associated with breast density. In our study, we hypothesized that the attention head with activations, typically for pixels numbering fewer than 50 but not exclusively, is likely representative of breast tissue. This hypothesis was formulated by analyzing the maximum number of pixel counts in each of the attention heads, which is less than 50 on 10\% of the training dataset. Based on this analysis, the sixth attention head was selected which mostly concentrates on the dense breast tissue. We then convert the attention visualization into binary masks using thresholding. Based on the binary mask, erasing is employed for the surrounding regions around the weakly localized dense tissues, a strategy termed the AGE data augmentation during transfer learning. This can be visually observed in Figure \ref{methodology}. AGE aims to enforce the network to focus more on the tissues than the background regions in the image. We experimented with performing the AGE augmentation during the downstream transfer learning task with various probability values (\emph{P}) of 0.2, 0.4, 0.6, and 0.8. The probability value indicates the likelihood of an image during training undergoing the augmentation. We added a classification layer on top of the DINO pretrained DeiT-small ViT for the four-class classification task. All experiments were trained for 50 epochs with a batch size of 8 and early stopping. A weighted binary cross-entropy loss function was optimized using a standard Adam optimizer with a learning rate of $5e^{-6}$ and a weight decay of $1e^{-4}$. We utilized a standard set of data augmentations as reported in \cite{panambur2022effect}.

\section{Results and Discussion}
\label{results}

\begin{table}[t]
\caption{Comparison of RE and AGE at different probabilities using Macro F1-scores. The best results are in \textbf{bold}.}
\label{table_results}
\begin{tabular*}{0.8\textwidth}{@{\extracolsep\fill}ccc}
\hline
Random Probability  & Random Erasing & Attention-Guided Erasing (Ours) \\
\hline
\emph{P} = 0.2 & 0.5691 (0.020) & 0.5774 (0.024) \\
\emph{P} = 0.4 & 0.5617 (0.034) & 0.5731 (0.017) \\
\emph{P} = 0.6 & 0.5522 (0.014) & \textbf{0.5910 (0.017)} \\
\emph{P} = 0.8 & 0.5503 (0.009) & 0.5747 (0.025) \\
\hline
\end{tabular*}
\end{table}

Table \ref{tab:scores} presents the classification performance using the mean macro F1-scores on the test dataset computed over five runs. We use the macro F1-score reported in \cite{soat} for comparison, where the authors achieve scores of 0.504 and 0.5525 using single-view ResNet-34 and EfficientNet-B2, respectively. In comparison, DINO-based transfer learning without any erasing augmentation achieves a mean F1-score of 0.5594. We then experiment by adding RE and AGE augmentations ($\textit{p}=0.2, 0.4, 0.6$ and $0.8$)~\cite{zhong2020random}. We observe an increase of almost 1\% with a score of 0.5691 when using the RE augmentation with a $\textit{P}$ value of 0.2. With the motivation to keep the class-specific features intact while erasing the background regions, we use our proposed AGE augmentation with various random probabilities. We observe a significant gain of more than 3.5\% in mean macro F1-score while using AGE compared to not using the AGE augmentation. We achieve a mean macro F1-score of 0.5910 in comparison to 0.5594 (No erasing) and 0.5691 (RE). This increase in gain suggests that AGE provides a better diversity of transformations during training, resulting in a more robust classification backbone. Moreover, we observe increased stability in the reported standard deviations (Table \ref{tab:scores}), indicating a regularization effect while using the proposed AGE augmentation. We also employ a two-tailed unpaired t-test to assess the statistical significance of the classification performances. We achieve a \emph{p}-value of \emph{p}<0.0001 emphasizing the statistical significance of our AGE approach. Table \ref{table_results} further depicts the mean macro F1 scores for each of the probability values for RE and AGE augmentations. It can be seen that the AGE augmentation consistently outperforms the classification performance of models trained without erasing and with RE. This is indicative of the fact that only the breast tissue is the class representative feature for density estimation, and the background regions of the breast can impact the classification performance. For example, the presence of pectoralis muscle and dense skin folds in MLO views, and the presence of suspicious regions of interest near the skin surface can potentially result in difficulty in the classification of breast density, especially in the case of high-class imbalance observed due to the higher percentage of dense breasts in the Vietnamese cohort. In the case of RE, we observe consistent performance drops, potentially due to erasing the class-relevant features in the majority of training samples with increasing probabilities. In future work, we aim to perform a comprehensive ablation study to investigate the impact of various data augmentation strategies in combination with AGE for breast density estimation. Furthermore, we plan to use AGE for other medical imaging modalities to check its generalization capabilities to more complex classification problems.

\begin{disclaimer}
The methods described in this paper are currently not available for commercial use, and there is no assurance of their future availability.
\end{disclaimer}

\tiny
\printbibliography

@article{globocon,
  author={Sung, Hyuna and Ferlay, Jacques and Siegel, Rebecca L and Laversanne, Mathieu and Soerjomataram, Isabelle and Jemal, Ahmedin and Bray, Freddie},
  journal={CA Cancer J Clin},
  number={3},
  pages={209--249},
  title={Global cancer statistics 2020: GLOBOCAN estimates of incidence and mortality worldwide for 36 cancers in 185 countries},
  volume={71},
  year={2021},
  publisher={Wiley Online Library}
}

@article{trieu2017risk,
  title={Risk factors of female breast cancer in Vietnam: a case-control study},
  author={Trieu, Phuong Dung Yun and Mello-Thoms, Claudia and Peat, Jennifer K and Do, Thuan Doan and Brennan, Patrick C},
  journal={Cancer research and treatment: Official journal of Korean Cancer Association},
  volume={49},
  number={4},
  pages={990--1000},
  year={2017},
  publisher={Korean Cancer Association}
}

@inproceedings{wang2014breast,
  title={Breast density and breast cancer risk: a practical review},
  author={Wang, Amy T and Vachon, Celine M and Brandt, Kathleen R and Ghosh, Karthik},
  booktitle={Mayo Clinic Proceedings},
  volume={89},
  number={4},
  pages={548--557},
  year={2014},
  organization={Elsevier}
}

@article{brandt2016comparison,
  title={Comparison of clinical and automated breast density measurements: implications for risk prediction and supplemental screening},
  author={Brandt, Kathleen R and Scott, Christopher G and Ma, Lin and Mahmoudzadeh, Amir P and Jensen, Matthew R and Whaley, Dana H and Wu, Fang Fang and Malkov, Serghei and Hruska, Carrie B and Norman, Aaron D and others},
  journal={Radiology},
  volume={279},
  number={3},
  pages={710--719},
  year={2016},
  publisher={Radiological Society of North America}
}

@article{gardezi2019breast,
  title={Breast cancer detection and diagnosis using mammographic data: Systematic review},
  author={Gardezi, Syed Jamal Safdar and Elazab, Ahmed and Lei, Baiying and Wang, Tianfu},
  journal={Journal of medical Internet research},
  volume={21},
  number={7},
  pages={e14464},
  year={2019},
  publisher={JMIR Publications Toronto, Canada}
}

@INPROCEEDINGS{soat,
  author={Nguyen, Huyen T. X. and Tran, Sam B. and Nguyen, Dung B. and Pham, Hieu H. and Nguyen, Ha Q.},
  booktitle={2022 44th Annual International Conference of the IEEE Engineering in Medicine \& Biology Society (EMBC)}, 
  title={A novel multi-view deep learning approach for BI-RADS and density assessment of mammograms}, 
  year={2022},
  volume={},
  number={},
  pages={2144-2148},
  doi={10.1109/EMBC48229.2022.9871564}}

@article{maghsoudi2021deep,
  title={Deep-LIBRA: An artificial-intelligence method for robust quantification of breast density with independent validation in breast cancer risk assessment},
  author={Maghsoudi, Omid Haji and Gastounioti, Aimilia and Scott, Christopher and Pantalone, Lauren and Wu, Fang-Fang and Cohen, Eric A and Winham, Stacey and Conant, Emily F and Vachon, Celine and Kontos, Despina},
  journal={Medical image analysis},
  volume={73},
  pages={102138},
  year={2021},
  publisher={Elsevier}
}

@inproceedings{panambur2022effect,
  title={Effect of Random Histogram Equalization on Breast Calcification Analysis Using Deep Learning},
  author={Panambur, Adarsh Bhandary and Madhu, Prathmesh and Maier, Andreas},
  booktitle={Bildverarbeitung f{\"u}r die Medizin 2022: Proceedings, German Workshop on Medical Image Computing, Heidelberg, June 26-28, 2022},
  pages={173--178},
  year={2022},
  organization={Springer}
}

@inproceedings{zhong2020random,
  title={Random Erasing Data Augmentation},
  author={Zhong, Zhun and Zheng, Liang and Kang, Guoliang and Li, Shaozi and Yang, Yi},
  booktitle={Proceedings of the AAAI Conference on Artificial Intelligence},
  volume={34},
  number={07},
  pages={13001-13008},
  year={2020}
}

@article{miao2022prior,
  title={Prior Knowledge-Guided Attention in Self-Supervised Vision Transformers},
  author={Miao, Kevin and Gokul, Akash and Singh, Raghav and Petryk, Suzanne and Gonzalez, Joseph and Keutzer, Kurt and Darrell, Trevor},
  journal={arXiv e-prints},
  pages={arXiv--2209},
  year={2022}
}

@inproceedings{dino,
  title={Emerging Properties in Self-Supervised Vision Transformers},
  author={Caron, Mathilde and Touvron, Hugo and Misra, Ishan and J\'egou, Herv\'e  and Mairal, Julien and Bojanowski, Piotr and Joulin, Armand},
  booktitle={Proceedings of the International Conference on Computer Vision (ICCV)},
  year={2021},
  url={https://github.com/facebookresearch/dino}
}

@incollection{birads,
    author = {Sickles, E. A. and D'Orsi, C. J. and Bassett, L. W. and others},
    booktitle = {ACR BI-RADS$^{\circledR}$ Atlas, Breast Imaging Reporting and Data System},
    title = {ACR BI-RADS$^{\circledR}$ Mammography},
    pages={121--140},
    year = {2013},
    publisher = {Reston, VA, American College of Radiology}
}

@article{vindr,
  title={VinDr-Mammo: A large-scale benchmark dataset for computer-aided diagnosis in full-field digital mammography},
  author={Nguyen, Hieu T and Nguyen, Ha Q and Pham, Hieu H and Lam, Khanh and Le, Linh T and Dao, Minh and Vu, Van},
  journal={Scientific Data},
  volume={10},
  number={1},
  pages={277},
  year={2023},
  publisher={Nature Publishing Group UK London}
}

\end{document}